\documentclass[aps,prb,twocolumn,showpacs,amssymb,floatfix]{revtex4-1}
\usepackage{amsmath,graphics,epsfig}
\usepackage{mathrsfs}
\usepackage{bm}
\usepackage{color}

\newcommand{\tx}{\text}
\newcommand{\ti}{\textit}

\newcommand{\la}{\langle}

\newcommand{\nn}{\nonumber}

\newcommand{\ra}{\rangle}

\newcommand{\alp}{\alpha}

\newcommand{\eps}{\epsilon}
\newcommand{\gm}{\gamma}

\newcommand{\sg}{\sigma}

\newcommand{\etal}{{\em et al.~}}
\newcommand{\ie}{{i.e.,~}}

\begin{document}
\title{Intraband and interband spin-orbit torques in non-centrosymmetric ferromagnets}
\author{Hang Li$^{1}$}\email[Both authors contributed equally to this work.]{}
\author{H. Gao$^{2,6}$}\email[Both authors contributed equally to this work.]{}
\author{Liviu P. Z\^arbo$^{3,4}$, K. V\'yborn\'y$^{4}$}\email{vybornyk@fzu.cz}
\author{Xuhui Wang$^{1}$, Ion Garate$^{5}$, \\ Fatih Do\v{g}an$^{1}$, A. \v{C}ejchan$^{4}$, Jairo Sinova$^{2,4,6}$}
\author{T. Jungwirth$^{4,7}$}
\author{Aur\'elien Manchon$^{1}$}\email{aurelien.manchon@kaust.edu.sa}
\affiliation{$^{1}$King Abdullah
University of Science and Technology (KAUST), Physical Science and
Engineering Division, Thuwal 23955-6900, Saudi Arabia,\\$^{2}$Department of Physics, Texas A\&M University, College Station, Texas 77843-4242, USA\\$^{3}$Molecular and Biomolecular Physics Department, National Institute for Research and Development of Isotopic and Molecular Technologies, RO-400293 Cluj-Napoca, Romania\\$^{4}$Institute of Physics ASCR, v.v.i., Cukrovarnick\'a 10, 162 53 Praha 6, Czech Republic\\$^5$D\'epartement de Physique and Regroupement Qu\'eb\'ecois sur les Mat\'eriaux de Pointe, Universit\'e de Sherbrooke, Sherbrooke, Qu\'ebec, Canada J1K 2R1.\\$^6$ Institute of Physics, Johannes Gutenberg Universit\"at, 55128 Mainz, Germany\\$^{7}$School of Physics and Astronomy, University of Nottingham, Nottingham NG7 2RD, United Kingdom}

\date{Jan07, 2015}

\begin{abstract}
 Intraband and interband contributions to the current-driven spin-orbit torque in magnetic materials lacking inversion symmetry are theoretically studied using Kubo formula. In addition to the current-driven field-like torque ${\bf T}_{\rm FL}= \tau_{\rm FL}{\bf m}\times{\bf u}_{\rm so}$ (${\bf u}_{\rm so}$ being a unit vector determined by the symmetry of the spin-orbit coupling), we explore the intrinsic contribution arising from impurity-independent interband transitions and producing an anti-damping-like torque of the form ${\bf T}_{\rm DL}= \tau_{\rm DL}{\bf m}\times({\bf u}_{\rm so}\times{\bf m})$. Analytical expressions are obtained in the model case of a magnetic Rashba two-dimensional electron gas, while numerical calculations have been performed on a dilute magnetic semiconductor (Ga,Mn)As modeled by the Kohn-Luttinger Hamiltonian exchanged coupled to the Mn moments. Parametric dependences of the different torque components and similarities to the analytical results of the Rashba two-dimensional electron gas in the weak disorder limit are described.
\end{abstract}

\pacs{72.25.Dc,72.20.My,75.50.Pp}
\maketitle

\section{\label{sec:intro}Introduction}
Magnetization dynamics driven electrically by spin-polarized currents through the spin transfer torque\cite{slonczewski-jmmm-1996,berger-prb-1996,ralph-jmmm-2008} has attracted considerable attention due to its applications in memory and logic spintronic devices.\cite{Chappert-nm-2007,Brataas-nm-2012} An alternative mechanism, the spin-orbit torque (SOT), has been recently proposed as a means to control the magnetization of a single ferromagnetic\cite{bernevig,manchon-prb,Obata-prb-2008} or even antiferromagnetic\cite{Zelezny:2014_a} layer without the need of an external spin-polarizer. The SOT arises from the interaction between the nonequilibrium spin density of carriers and the local magnetization. The non-equilibrium spin density results from the transfer of angular momentum between the spin and orbital degrees of freedom of the carriers.\cite{bernevig,manchon-prb,Obata-prb-2008,garate-prb-2009,pesin2012,bijl2012,kim2012,wang-manchon-2012,Hang-apl-2013} The SOT requires magnetic structures with strong spin-orbit coupling and inversion symmetry breaking. Initially observed in epilayers of (Ga,Mn)As dilute magnetic semiconductors (DMSs) with bulk inversion asymmetry in their strained zinc-blende crystal,\cite{chernyshov-nph-2009,endo-apl-2010,fang-nanotech-2011} this effect was soon widely confirmed in metallic bilayers with structural inversion symmetry breaking.\cite{miron1,miron2,liu,kim,garello,fan,jamali,mellnik} In general, the SOT observed experimentally possesses two components, a field-like torque ${\bf T}_{\rm FL}= \tau_{\rm FL}{\bf m}\times{\bf u}_{\rm so}$ {\em odd} in the magnetization direction ${\bf m}$ and an \hbox{anti-damping-like} torque\cite{note2} ${\bf T}_{\rm DL}= \tau_{\rm DL}{\bf m}\times({\bf u}_{\rm so}\times{\bf m})$ {\em even} in {\bf m}. Here, ${\bf u}_{\rm so}$ is a unit vector determined by the symmetry of the structure and the current direction,\cite{note1} and $\tau_{\rm FL}$ and $\tau_{\rm DL}$ are the magnitudes of the field-like and anti-damping-like torque, respectively. These torques are also commonly referred to as the {\em out-of-plane} and {\em in-plane} torques, respectively, with respect to the (${\bf m},{\bf u}_{\rm so}$) plane. The direction of the field-like (out-of-plane) and anti-damping-like (in-plane) torques and their detailed angular dependence\cite{note1} depend on the crystal structure, while their magnitude has been shown to strongly depend on the materials considered. \cite{kim,garello,fan,jamali,mellnik}\par

Two main mechanisms have been invoked to explain the origin of the
current-driven torques in non-centrosymmetric ferromagnets. In the
first scenario, the lack of inversion symmetry enables the 
inverse spin galvanic effect\cite{isge-reviews} (ISGE), 
i.e. flowing current
directly produces  a nonequilibrium spin density $\delta{\bf S}$
locally, whose direction is determined by the symmetry of the
spin-orbit coupling. 
Recently, it has been proposed that in non-centrosymmetric magnetic
materials this nonequilibrium spin density may exert a torque on the
magnetization\cite{bernevig,manchon-prb,Obata-prb-2008,garate-prb-2009} 
${\bf T}=(2J_{\rm ex}/\hbar\gamma N_m){\bf m}\times\delta{\bf S}$. Here,
$\gamma$ is the gyromagnetic ratio, $N_m$ the density of magnetic
moments and $J_{\rm ex}$ the exchange coupling (having the dimension
of energy)  between the itinerant electron spins and the local
magnetization ${\bf M}=M_s{\bf m}$ which, in this article, is assumed
to arise solely from localized magnetic moments $\mu$ so that the
saturated magnetization $M_s=\mu N_m$.
This is the essence of the ISGE-induced SOT. Alternatively, in
ferromagnets adjacent to a heavy metal, it has also been proposed that
the spin Hall effect (SHE) present in the heavy metal may inject a
spin-polarized current into the adjacent ferromagnet, exerting a
spin-transfer torque (STT) on the magnetization.\cite{liu,haneyboltz,miron2}

A current debate aims at identifying the interplay between these different mechanisms and their impact in terms of current-driven spin torque. In the simplest physical picture, SHE induces an anti-damping-like STT, while the SOT reduces to a field-like torque generated by ISGE.\cite{haneyboltz} However, it has been recently proposed that the incomplete absorption of the SHE-induced spin current by the ferromagnet (or, equivalently, the non-vanishing imaginary part of the interfacial spin mixing conductance) may result in a field-like STT component.\cite{haneyboltz} Similarly, in the context of ISGE-induced SOT, recent theories have suggested that spin relaxation and dephasing may also lead to a correction in the SOT in the form of a anti-damping-like component.\cite{pesin2012,bijl2012,kim2012,wang-manchon-2012} In Refs. \onlinecite{bijl2012} and \onlinecite{kim2012}, the anti-damping-like SOT term arises from the electron scattering-induced spin relaxation. In Ref.~\onlinecite{haneyboltz}, the semiclassical diffusion formalism was used, whereas in Refs. \onlinecite{pesin2012} and \onlinecite{wang-manchon-2012}, the anti-damping-like SOT is obtained within a quantum kinetic formalism. It is ascribed to spin-dependent carrier lifetimes \cite{pesin2012} or to a term arising from the weak-diffusion limit, which in the leading order is proportional to a constant carrier lifetime.\cite{wang-manchon-2012} 

Intriguing material-dependence of the SOTs has been unravelled in various experiments keeping the debate on the origin of these components open.\cite{kim,garello,fan,jamali,mellnik} The difficulty in determining the physical origin of the torques partly lies in the complexity of the ultrathin bilayer considered, involving both bulk and interfacial transport in the current-in-plane configuration. First principle calculations have indeed pointed out the significant sensitivity of the torques to the nature of the interfaces.\cite{haneydft} \par

In a recent publication, Kurebayashi et al.\cite{Kurebayashi} investigated the SOT in a bulk DMS. They observed a large anti-damping-like torque that is not ascribed to the SHE since no adjacent spin-orbit coupled paramagnet is present. It was then proposed that such a torque has a scattering-independent origin in the Berry curvature of the band structure, in a similar spirit as the intrinsic SHE was introduced about ten years ago.\cite{sinova2004,Murakami:2003_a}\par

In this paper, we present a systematic theoretical study of SOTs
arising from the ISGE and Berry curvature mechanisms in a spin-independent 
relaxation time approximation.
We focus our attention on the current-driven spin-orbit field (called the SOT field), ${\bf h}^{\rm so}$, producing the spin-orbit torque ${\bf T}={\bf M}\times{\bf h}^{\rm so}$. These SOT fields have an {\em in-plane} component of the ISGE origin\cite{note1} ${\bf h}_{\|}^{\rm so}=\tau_{\rm FL}{\bf u}_{\rm so}$ [i.e. lying in the (${\bf m},{\bf u}_{\rm so}$) plane and producing an out-of-plane torque] and also an intrinsic contribution arising from interband transitions.
The latter\cite{Kurebayashi} 
produces an {\em out-of-plane} field of the form ${\bf h}_{\bot}^{\rm so}=\tau_{\rm DL}{\bf u}_{\rm so}\times{\bf m}$ [i.e. lying perpendicular to the (${\bf m},{\bf u}_{\rm so}$) plane]. Analytical expressions are obtained in the model case of a magnetic Rashba two-dimensional electron gas (2DEG), while numerical calculations are performed on DMSs described by the kinetic-exchange Kohn-Luttinger Hamiltonian.\cite{dms-rmp} Parametric dependences of the different torque components and similarities to the analytical results of the Rashba two-dimensional electron gas in the weak disorder limit are described.


\section{\label{sec:model}Non-equilibrium spin density: intraband and Interband Contributions in Kubo formula}
In the present study, we start from a general single-particle Hamiltonian
\begin{align}
\hat{H}_{\tx{sys}}=\hat{H}_{\tx{0}}+\hat{H}_{\tx{SOC}}+\hat{H}_{\tx{ex}}+V_{\rm imp}({\bf r})-e{\bf E}\cdot\hat{\bf r},
\label{eq:total-Hamiltonian}
\end{align} 
where the first term includes the spin-independent kinetic and potential energies of the particle, the second term denotes the coupling between the carrier spin and its orbital angular momentum and the third one represents the interaction between the spin of the carrier and the magnetization of the ferromagnetic system. Below, we refer to these first three terms as to the unperturbed part of the Hamiltonian. The fourth term is the impurity potential and the fifth term is the electric field applied through the system. Impurities are treated within the constant relaxation time approximation while the electric field is treated within the framework of the linear response theory. As discussed below, this electric field has two distinct effects on the electronic system: (i) it modifies the carrier distribution function from its equilibrium Fermi-Dirac form and (ii) it distorts the carrier wave functions. The former leads to intraband ISGE contributions, while the latter is responsible for the interband (Berry curvature) contribution. To calculate the SOT field, we evaluate first the nonequilibrium spin density $\delta {\bf S}$ using the Kubo formula
\begin{align}
\delta{\bf S}=&\frac{e\hbar}{2\pi V}{\rm Re}\sum_{{\bf k},a,b} \la\psi_{{\bf k}a}|\hat{\bf{s}}|\psi_{{\bf k}b}\ra\la\psi_{{\bf k}b}|  {\bf E}\cdot \hat{\bf{v}}|~\psi_{{\bf k}a}\ra\nn\\
&\times [G^{R}_{{\bf k}a}G^{A}_{{\bf k}b}-G^{R}_{{\bf k}a}G^{R}_{{\bf k}b}],
\label{eq:TSP}
\end{align}
where $G^{R}_{{\bf k}a}=(G^{A}_{{\bf k}a})^{*}=1/(E_{F}-E_{{\bf k}a}+i\Gamma)$, $E_{F}$ is the Fermi energy, $E_{{\bf k}a}$ is the energy dispersion of band $a$, $V$ is the system volume and $\Gamma$ is the spectral broadening due to the finite lifetime of the particle in the presence of impurities. The Bloch state $|\psi_{{\bf k}a}\ra$ in band $a$ can be found by diagonalizing the unperturbed part of the Hamiltonian in Eq.~(\ref{eq:total-Hamiltonian}). This expression contains both intraband ($a=b$) and interband ($a\neq b$) contributions to the nonequilibrium spin density. Numerical results in Section \ref{sec:numerical results} are calculated with the above equation.

In order to understand the numerical results, Eq.~(\ref{eq:TSP})
can be rewritten\cite{note3} as $\delta{\bf S}=\delta{\bf S}^{\rm intra} +
\delta{\bf S}^{\rm inter}_{1}+\delta{\bf S}^{\rm inter}_{2}$
when weak impurity scattering (namely,
small spectral broadening, $\Gamma\rightarrow0$) is assumed. The three
contributions are
\begin{align}\label{eq:kubo1}
\delta{\bf S}^{\rm intra}&=\frac{1}{V}\frac{e\hbar}{2\Gamma}\sum_{{\bf k},a}
\la\psi_{{\bf k}a}|\hat{\bf{s}}|\psi_{{\bf k}a}\ra\la\psi_{{\bf k}a}|{\bf E}\cdot \hat{\bf{v}}|~\psi_{{\bf k}a}\ra\nn\\
&\times \delta(E_{{\bf k}a}-E_F),\\
\delta{\bf S}^{\rm inter}_{1}=&-\frac{e\hbar}{V}\sum_{{\bf k},a\not= b}
2{\rm Re}[\la\psi_{a\bm{k}}|\hat{\bf{s}}|\psi_{b\bm{k}}\ra \la\psi_{b\bm{k}}|{\bf E}\cdot \hat{\bf{v}}|\psi_{a\bm{k}}\ra]\nonumber\\
& \times \frac{\Gamma(E_{{\bf k}a}-E_{{\bf k}b})}{[(E_{{\bf k}a}-E_{{\bf k}b})^2+\Gamma^2]^2}(f_{{\bf k}a}-f_{{\bf k}b}).\label{eq:kubo2}\\
\delta{\bf S}^{\rm inter}_{2}=&-\frac{e\hbar}{V}\sum_{{\bf k},a\not= b}
{\rm Im}[\la\psi_{{\bf k}a}|\hat{\bf{s}}|\psi_{{\bf k}b}\ra \la\psi_{{\bf k}b}|{\bf E}\cdot \hat{\bf{v}}|\psi_{{\bf k}a}\ra]\nonumber\\
& \times \frac{\Gamma^2-(E_{{\bf k}a}-E_{{\bf k}b})^2}{[(E_{{\bf k}a}-E_{{\bf k}b})^2+\Gamma^2]^2}(f_{{\bf k}a}-f_{{\bf k}b}).
\label{eq:kubo3}
\end{align}
The first term, Eq.~(\ref{eq:kubo1}), is the intraband ($a=b$)
contribution arising from the perturbation of the carrier
distribution function by the electric field. It is proportional to the momentum
scattering time ($\tau= \hbar/2\Gamma$) and is therefore an
{\em extrinsic} contribution to the nonequilibrium spin density
(i.e. it is impurity-dependent). The second and third terms are interband
($a\neq b$) contributions arising from the perturbation of the carrier
wave functions by the electric field. The second term, Eq.~(\ref{eq:kubo2}), 
is inversely proportional to the scattering time, i.e. it vanishes in 
the clean limit. The third term, Eq.~(\ref{eq:kubo3}),
is independent of the scattering in the weak scattering limit 
($E_{{\bf k}a}-E_{{\bf k}b}\gg\Gamma$), i.e. 
is an {\em intrinsic} contribution to
the nonequilibrium spin density. The formalism described above is 
the established linear response theory of a translationally
invariant system and has been exploited, for instance, in the context
of the spin Hall\cite{Sinova:2014_a} and anomalous Hall 
effect.\cite{rmp} Nevertheless, the distinction between these different 
contributions is particularly important in the case of the SOT since 
these terms give rise to different symmetries of the torque.\par

\begin{figure}[h!]
\begin{center}
\includegraphics[scale=0.4]{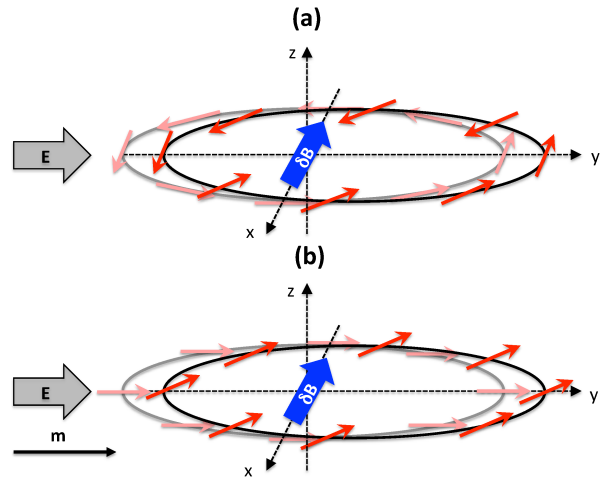}
\end{center}
\caption{(Color online) (a) Fermi surface of a non-magnetic Rashba 2DEG: under the application of an external electric field (thick grey arrow), a non-equilibrium field $\delta{\bf B}$ is produced (thick blue arrow) that distorts the spin direction out of the plane (red arrows). After averaging, the spin density vanishes. (b) Fermi surface of a magnetic Rashba 2DEG in the strong ferromagnetic regime: in this case, since the spin directions (pink arrows) are initially mostly aligned along the magnetization {\bf m} (black arrow), the resulting non-equilibrium spin density (red arrows) does not vanish and is aligned along ${\bf z}$.}
\label{Fig0}
\end{figure}

The concept of intrinsic SOT is illustrated in Fig.~\ref{Fig0} (see also the discussion in Ref. \onlinecite{Kurebayashi}). Figure~\ref{Fig0}(a) represents the Fermi surface of a non-magnetic Rashba 2DEG under the application of an external electric field ${\bf E}$. At equilibrium (${\bf E}=0$) the spin direction (pink arrows) is tangential to the Fermi surface (grey circle) at all k-points, and the total spin density vanishes. Applying the electric field accelerates the electrons on the Fermi surface and they feel a modified spin-orbit field $\delta{\bf B}\propto {\bf z}\times\dot{\bf p}= -{\bf z}\times e{\bf E}$ (thick blue arrow) around which the spin momenta (red arrows) start to precess. For a non-magnetic 2DEG, the resulting spin density vanishes while a non-zero transverse spin-current is generated by this mechanism. This is the origin of the intrinsic SHE.

Now, let us consider the case of a magnetic Rashba 2DEG in the strong ferromagnetic limit [Fig.~\ref{Fig0}(b)]. At equilibrium, the spin momenta (pink arrows) are approximately aligned along the magnetization direction ${\bf m}$ (thin black arrow) for all k-points of the Fermi surface (grey circle). Under the application of an external electric field, the spin momenta (red arrows) precess around $\delta{\bf B}$ (thick blue arrow) resulting in a non-vanishing spin density. Following the convention adopted in Fig.~\ref{Fig0}(b), the electric field and equilibrium magnetization are along ${\bf y}$, the displacement of the Fermi surface produces a non-equilibrium spin-orbit field $\delta {\bf B}$ along ${\bf x}$ and the spin precession around $\delta{\bf B}$ produces a spin density aligned along ${\bf z}$. The latter results in an additional torque that has a disorder-independent origin.\cite{Kurebayashi} This simple picture can be extended to more complex spin-orbit coupling situation and only requires inversion symmetry breaking in the system.

\section{\label{sec:2deg}Two-dimensional Rashba ferromagnet}
We first apply this formalism to a ferromagnetic 2DEG in the presence of Rashba spin-orbit coupling. \cite{rashba-1984} This system is the prototypical free-electron model for SOTs in ultra thin ferromagnets embedded between two asymmetric interfaces.\cite{manchon-prb,Obata-prb-2008,garate-prb-2009} Although the actual band structure of magnetic bilayers such as Pt/Co is complex, recent first principle calculations indicate that this simple Rashba model qualitatively captures most of the relevant physics at these interfaces.\cite{haneydft} This section is therefore developed mostly for pedagogical purposes in order to make the dependence on various parameters explicit. The unperturbed Hamiltonian in Eq.~(\ref{eq:total-Hamiltonian}) can be rewritten as: 
\begin{align}
\hat{H}_{\tx{2DEG}}=\frac{\hbar^2k^{2}}{2m^*}-\alp \hat{\bm\sigma}\cdot({\bf z}\times{\bf k})+J_{\rm ex}{\bf{m}} \cdot \hat{\bm\sg}.\label{eq:total-hamiltonian2}
\end{align}
where ${\bf k}=(k_x,k_y,0)=k(\cos\varphi_k,\sin\varphi_k,0)$, $\alp$ is the Rashba parameter and the magnetization direction is ${\bf m}=(\cos \varphi \sin \theta, \sin \varphi \sin \theta, \cos \theta )$. By diagonalizing Eq.~(\ref{eq:total-hamiltonian2}), the eigenvalues and eigenvectors of itinerant electrons are
\begin{eqnarray}
&&E_{\bf{k}\pm}=\frac{\hbar^2k^{2}}{2m^*}\pm \Delta_{\bf k},\\
&& \Delta_{\bf k}=\sqrt{J_{\rm ex}^2+\alp^ 2k^2+2\alp kJ_{\rm ex}\sin (\varphi-\varphi_{k}) \sin \theta},\\
 &&|\bf{k},+\rangle=      
\left(                 
  \begin{array}{c} 
    e^{i\gamma_{\bf{k}}}\cos \frac{\chi_{k}}{2} \\  
    \sin \frac{\chi_{k}}{2}\\  
  \end{array}
\right) ; 
|\bf{k},-\rangle=        
\left(                 
  \begin{array}{c} 
   -e^{i\gamma_{\bf{k}}}\sin \frac{\chi_{k}}{2} \\  
    \cos \frac{\chi_{k}}{2}\\  
  \end{array}
\right)       
\end{eqnarray}
where we have $\cos\chi_{k}=J_{\rm ex}\cos\theta/\Delta_{\bf k}$ and $\tan \gamma_{\bf{k}}=\frac{\alp k\cos \varphi_{k}+J_{\rm ex} \sin \varphi \sin \theta}{\alp k\sin \varphi_{k}-J_{\rm ex} \cos \varphi \sin \theta}$.
The velocity operator is given by $\hat{\bf v}=\frac{\hbar}{m^*}{\bf k}+\frac{\alp}{\hbar} {\bf z} \times \hat{\bm \sg}$. Exploiting Eqs.~(\ref{eq:kubo1})-(\ref{eq:kubo3}) in the weak exchange limit ($\alp k_{\rm F}\gg J_{\rm ex}\gg \Gamma$), the nonequilibrium spin density reads
\begin{align}\label{eq:intraw}
\delta {\bf S}^{\rm intra}&=\frac{1}{4\pi}\frac{\alp m^{*}}{\hbar^{2}\Gamma}({\bf z} \times e{\bf E})\\\label{eq:inter1w}
\delta {\bf S}^{\rm inter}_1&=-\frac{1}{8\pi}\frac{\Gamma}{\alp E_{F}}({\bf z} \times e{\bf E})\\\label{eq:inter2w}
\delta {\bf S}^{\rm inter}_2&=\frac{1}{4\pi}\frac{J_{\rm ex}}{\alpha E_F}({\bf m}\cdot{\bf z}) e{\bf E}
\end{align}
and in the strong exchange limit ($J_{\rm ex} \gg \alp k_{\rm F}\gg \Gamma$),
\begin{align}\label{eq:intras}
\delta {\bf S}^{\rm intra}&=\frac{1}{2\pi}\frac{\alp m^{*}}{\hbar^{2}\Gamma}{\bf m}\times[({\bf z} \times e{\bf E})\times{\bf m}]\\\label{eq:inter1s}
\delta {\bf S}^{\rm inter}_1&=-\frac{1}{2\pi}\frac{\alp m^{*}\Gamma}{\hbar^{2} J_{\rm ex}^2}{\bf m}\times[({\bf z} \times e{\bf E})\times{\bf m}]\\\label{eq:inter2s}
\delta {\bf S}^{\rm inter}_2&=-\frac{1}{2\pi}\frac{\alp m^{*}}{\hbar^{2} J_{\rm ex}}{\bf m}\times({\bf z} \times e{\bf E})
\end{align}
In summary, the SOT field defined as 
${\bf h}=2J_{\rm ex} \delta{\bf S}/\gamma\hbar N_m$, takes on the following
form in the two limits:
\begin{align}\label{eq:sofw}
J_{\rm ex}\ll\alp k_{\rm F}\!:\ {\bf h}=&\frac{J_{\rm ex} \alp m^{*}}{2\pi\gamma N_m\hbar^{3}\Gamma}\left(1-\frac{\Gamma^2}{\alp^2k_{\rm F}^2}\right)({\bf z} \times e{\bf E})\nonumber\\
&+\frac{J_{\rm ex}^2}{2\pi\gamma N_m\hbar\alpha E_F}({\bf m}\cdot{\bf z}) e{\bf E},\\
J_{\rm ex}\gg\alp k_{\rm F}\!:\ {\bf h}=&\frac{J_{\rm ex}\alp m^{*}}{\pi\gamma N_m\hbar^{3}\Gamma}\left(1-\frac{\Gamma^2}{J_{\rm ex}^2}\right){\bf m}\times[({\bf z} \times e{\bf E})\times{\bf m}]\nonumber\\
&+\frac{\alp m^{*}}{\pi\gamma N_m\hbar^{3}}{\bf m}\times({\bf z} \times e{\bf E})\label{eq:sofs}
\end{align}
Three important facts ought to be pointed out. First, the extrinsic contributions (either intra- or interband) both give rise to an in-plane SOT field [even in magnetization direction, lying in the (${\bf m},{\bf z}\times{\bf E}$) plane]. The resulting extrinsic torque is then out-of-plane and odd in magnetization direction. Second, the intrinsic contribution [second term in Eqs.~(\ref{eq:sofw}) and (\ref{eq:sofs})] only produces a SOT field odd in magnetization direction. It lies perpendicular to the (${\bf m},{\bf z}\times{\bf E}$) plane in the strong exchange limit, see Eq.~(\ref{eq:sofs}). This term is {\em independent} of the exchange $J_{\rm ex}$, in sharp contrast with the ISGE-induced SOT field, while in the weak exchange limit, Eq.~(\ref{eq:sofw}), it is second order in exchange and proportional to $m_z e {\bf E}$. The resulting intrinsic torque is in-plane in the strong exchange limit and even in magnetization direction. As will be seen in the next section, the parameters dependence displayed in Eqs.~(\ref{eq:sofw}) and (\ref{eq:sofs}) is not restricted to the simple case of the Rashba model. Third, notice that in the strong exchange limit the ratio of the anti-damping-like to the field like torque is $\approx \Gamma/J_{\rm ex}$. This dependence is the inverse of what was found in Refs. \onlinecite{pesin2012} and \onlinecite{wang-manchon-2012} in the diffusive limit and ignoring the interband scattering, where the ratio between the two torques is governed by $J_{\rm ex}/\Gamma$. A corrective anti-damping-like torque proportional to $\Gamma_{\rm sf}/J_{\rm ex}$ is also obtained when considering a finite spin-flip relaxation time $\tau_{\rm sf}=\hbar/\Gamma_{\rm sf}$.\cite{wang-manchon-2012,pesin2012}\par

In the case of the anomalous Hall effect, related and better explored
transport phenomenon, the intrinsic contribution dominates over the extrinsic contributions in the strong scattering limit.\cite{onoda,rmp} As a consequence, one is tempted to anticipate that the intrinsic contribution to the SOT discussed presently becomes important when strong momentum scattering is present (such as in disordered Pt/Co interfaces for example) and dominates over the corrections found in Refs.~\onlinecite{wang-manchon-2012,pesin2012} in this limit. Nevertheless, these different contributions have been derived in different limits --- i.e. strong\cite{wang-manchon-2012,pesin2012} versus weak scattering (this work) --- and should to be treated on equal footing for a rigorous comparison (e.g., see Ref. \onlinecite{onoda}). Such a comprehensive model is beyond the scope of the present work.

\section{Dilute Magnetic Semiconductors}
\subsection{Method}
We now extend the previous results beyond the simple ferromagnetic 2DEG model with Rashba spin-orbit coupling. We consider a bulk three-dimensional DMS, such as (Ga,Mn)As, with a homogeneous magnetization. In order to model the SOT field of (Ga,Mn)As, we adopt a Hamiltonian including a mean-field exchange coupling between the hole spin ($\hat{\bf{J}}$) and the localized ($d$-electron) magnetic moment $\mu S_{a}{\bf m}$ of ionized $\tx{Mn}^{2+}$ acceptors \cite{abolfath-prb-2001,jungwirth-apl-2002} and a four-band strained Kohn-Luttinger Hamiltonian.
The total Hamiltonian of the DMS reads 
\begin{align}
{\hat H}_{\tx{DMS}}={\hat H}_{\tx{L}}+{\hat H}_{\tx{strain}}+J_{\tx{pd}}N_{\tx{Mn}}S_{a}{\bf m}\cdot{\hat{\bf{J}}}
\label{eq:Ha_o}
\end{align}
where $J_{\tx{pd}}=55~\tx{meV}\cdot\tx{nm}^3$ is the antiferromagnetic 
coupling constant between hole and local moment spins for (Ga,Mn)As and ${S_{a}=5/2}$ is the localized Mn spin.
The hole spin operator is a $4\times 4$ matrix.\cite{abolfath-prb-2001} The 
concentration of the ordered local Mn$^{2+}$ moments $N_{\tx{Mn}}=4x/a^3$ 
is the product of $x$ that defines the doping by Mn$^{2+}$ ions and inverse 
volume per Ga atom ($a$ is the GaAs lattice constant). The Kohn-Luttinger 
Hamiltonian in Eq.~(\ref{eq:Ha_o}) is expressed as \cite{luttinger56}
\begin{align}
{\hat H}_{\tx{L}} = \frac{\hbar^2}{2m}
&\left[\gamma_1k^2{\hat{\rm I}}
-4\gamma_3{[k_{x}k_{y}\{\hat{\rm J}_{x},\hat{\rm J}_{y}}\}+c.p.]\right.\nn\\
&\left. -2\gamma_2[(\hat{\rm J}_{x}^2-\frac{1}{3}\hat{\rm{\bf J}}^2)k_{x}^2+c.p.]\right].
\label{eq:Ha5}
\end{align}
This Hamiltonian applies close to the $\Gamma$ point to centro-symmetric crystals with a diamond structure and strong spin-orbit coupling in the valence bands. The Luttinger parameters for GaAs are $(\gamma_1,\gamma_2,\gamma_3)=(6.98,2.06,2.93)$, $\hat{\rm{I}}$ is the $4\times4$ unity matrix, $\hat{\rm J}_{x}$, $\hat{\rm J}_{y}$, and $\hat{\rm J}_{z}$, are the angular momentum matrices for spin $\frac{3}{2}$. They follow the relation $\{\hat{\rm J}_{x},\hat{\rm J}_{y}\}=(\hat{\rm J}_{x}\hat{\rm J}_{\rm y}+\hat{\rm J}_{ y}\hat{\rm J}_{\rm x})/2$, and $c.p.$ denotes cyclic permutation of the preceding term. The first term denotes the kinetic energy of the holes. The second and third terms are associated with the spin-orbit coupling of the diamond crystal. In zinc-blende crystals, such as GaAs, bulk inversion asymmetry gives rise to the so-called cubic Dresselhaus spin-orbit coupling.\cite{cubic-dresselhaus} We neglect this term in the present study since there is no experimental indication that it contributes significantly to the SOT in (Ga,Mn)As. 

Hamiltonian ${\hat H}_{\tx{DMS}}$ should be understood as an effective model attempting to describe the current-driven SOT in (Ga,Mn)As rather than the complete description of the electronic structure in this material. In a cubic diamond crystal, $\gm_{2}\neq\gm_{3}$. When $\gm_{2}=\gm_{3}$, dispersions following from Eq.~(\ref{eq:Ha5}) become spherically symmetric and when the spin-orbit coupling is removed completely ($\gm_{1}=2.0,\gm_{2}=\gm_{3}=0$), Eq.~(\ref{eq:Ha5}) reduces to a parabolic model. The impact of these three degrees of approximation (parabolic model, spherical approximation and diamond crystal) on the SOT will be addressed in Section \ref{sec:fermi-surface}. \par

At this level of approximation the effective Hamiltonian, Eq.~(\ref{eq:Ha5}), does not break bulk inversion symmetry even though the actual crystal of the host GaAs does. Indeed, although the
full model of GaAs contains additional terms that are odd in ${\bf k}$ (see Tab.~6.2 in Ref.~\onlinecite{Band-Winkler}), it is experimentally established that the SOT in (Ga,Mn)As is
sensitive to the strain. We therefore assume, in line with experiments,\cite{fang-nanotech-2011} that the key inversion-breaking term is proportional to the strain. The strain Hamiltonian is given by
\begin{align}
\hat{H}_{\tx{strain}}&=C_{4}[\hat{\rm J}_{x}k_x(\eps_{yy}-\eps_{zz})+c.p.]\nonumber\\
&+C_{5}[(\hat{\rm J}_{x}k_y-\hat{\rm J}_{y}k_x)\eps_{xy}+c.p.]
\label{eq:Ha}
\end{align}
where $\eps_{ii}$ and $\eps_{ij}$ ($i\neq j$) are the diagonal and
non-diagonal elements of the strain tensor, respectively. We assume
$\eps_{xx}\equiv\eps_{yy}$ and $\eps_{xy}\equiv\eps_{yx}$. 

In Eq.~(\ref{eq:Ha}), the first term ($\propto C_4$) originates from the
lattice mismatch between the crystal structure of the substrate and
the one of (Ga,Mn)As, and produces a spin-orbit coupling with
Dresselhaus symmetry \cite{linear-dresselhaus} ($\propto \epsilon_{zz}$). The second term
($\propto C_5$) is the shear strain and possesses the symmetry of Rashba spin-orbit coupling \cite{rashba-1984} ($\propto \epsilon_{xy}$). Among the different terms linear in ${\bf k}$ and resulting from the inversion symmetry breaking (see Tab.~C.5 in Ref.~\onlinecite{Band-Winkler}), $\hat{H}_{\tx{strain}}$ is the only one that acts in the manifold of heavy and light-hole states. It is worth pointing out that we
consider here a large-enough system that allows us to disregard any
effects arising from boundaries and confinement. In the following, we
assume $C_{4}=C_{5}= 
0.5$~eV$\cdot$nm\cite{fang-nanotech-2011,chernyshov-nph-2009}
and consider the lattice mismatch strain (with
$\eps_{zz}\not=\eps_{yy}=\eps_{xx}=0$ and $\eps_{xy}=0$).
Physical presence of the shear strain ($\eps_{xy}\not=0$) in 
unpatterned (Ga,Mn)As
samples is below the detection limit,\cite{Kopecky:2011_a} yet it has been
introduced in previous theory studies to effectively model the in-plane uniaxial
anisotropy observed in experiments.\cite{Zemen:2009_a} 
Calculations with nonzero $\eps_{xy}$ are explicitly pointed out 
in the following. 

The SOT field ${\bf h}=2J_{\tx{pd}}\delta{\bf S}/\gamma\hbar$ 
is evaluated once the energies $E_{{\bf k}a}$ and eigenfunctions 
$|\psi_{{\bf k}a}\ra$ implied by the Hamiltonian in Eq.~(\ref{eq:Ha_o})
are numerically calculated and the current-driven spin density
$\delta{\bf S}$ is determined using Eq.~(\ref{eq:TSP}). In general,
the SOT field can be decomposed as
\begin{equation}\label{eq-16}
{\bf h}=h_{\rm m} {\bf m}+h_\|\hat{e}_{\parallel}+h_\bot\hat{e}_{\perp}
\end{equation}
where vectors $\hat{e}_{\parallel},\hat{e}_{\perp}$ have unit length,
$\hat{e}_\perp||{\bf m}\times{\bf u}$, $\hat{e}_\|=\hat{e}_\perp\times
{\bf m}$,
the subscript ``${\rm so}$'' has been removed for simplicity,
and the direction of ${\bf u}$ (whose length is also set
equal to one) should be chosen depending on the system. For example,
we find ${\bf u}||{\bf z}\times{\bf E}$ for the Rashba 2DEG. On the
other hand, ${\bf u}={\bf x}$ in (Ga,Mn)As with growth strain
($\propto \epsilon_{zz}$) as described by Eq.~(\ref{eq:Ha_o}) and
current flowing along the $[100]$ crystallographic direction. Our
results presented below always assume $\hat{e}_\perp$ pointing in the
positive ${\bf z}$ direction.

In the following, we disregard the component of the SOT field which is
parallel to the magnetization ($h_{\rm m}$) since it does 
not exert any torque on
it. The two remaining components in Eq.~(\ref{eq-16}) turn out to
produce, in (Ga,Mn)As, the anti-damping-like SOT in the case of
$h_\bot$ which is due to {\em intrinsic} interband mixing (of
impurity-independent origin) and a combination of anti-damping-like
and field-like {\em extrinsic} SOT in the case of $h_\|$ which depends through
$\Gamma$ on the disorder strength. The angular dependence of 
the two components, $h_{\|,\bot}$, reflects the details of the band
structure as discussed in Sec.~\ref{sec:fermi-surface}.

\subsection{\label{sec:numerical results}Numerical Results}

For all the calculations presented in this section, the electric field
$E=0.02$~V/nm is assumed to be applied along the $x$-axis and 
we varied the hole concentration
between 0.3~nm$^{-3}$ and 1~nm$^{-3}$. This corresponds, respectively,
to a Fermi energy between 200 and 450~meV. Except for
Sec.~\ref{sec:fermi-surface}, magnetization always lies along the
$y$-axis ($\varphi=90^\circ$).

\subsubsection{Intrinsic Versus Extrinsic Spin-Orbit Torques}

\begin{figure}[h!]
\begin{center}
\includegraphics[trim=0mm 0mm 0mm 0mm,clip,scale=0.4]{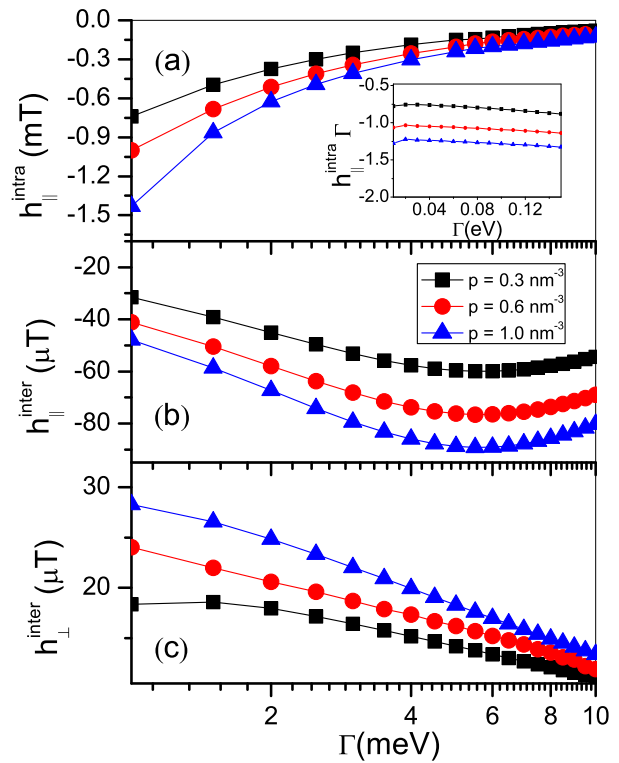}
\end{center}
\caption{(Color online) (a) Intraband and (b)-(c) interband
  contributions to the SOT field as a function of spectral
  broadening $\Gamma$ for otherwise typical (Ga,Mn)As sample (doping
  concentration $x=5\% $, lattice-mismatch strain $\eps_{zz}=-0.3\%$).
  Inset of panel (a) shows that $h_\|^{\rm intra}\propto 1/\Gamma$ holds 
  over a broad range of $\Gamma$. Only lattice-mismatch strain is considered,
  so that $\eps_{xy}=0$ in Eq.~(\ref{eq:Ha}).}
\label{fig6}
\end{figure}

We first investigate the impact of impurity scattering on the
intraband and interband contributions to the SOT fields. 
Figure~\ref{fig6} displays the SOT field as a function of the energy
broadening $\Gamma$ for different values of hole
concentrations. Although $\Gamma$ is of the order of hundreds of meV
in realistic (Ga,Mn)As, we choose $\Gamma<10$~meV so as to be able to
compare these results with the analytical ones obtained in
Sec.~\ref{sec:2deg} for the ferromagnetic Rashba 2DEG which are valid
in the small $\Gamma$ limit.

The intraband contribution to the SOT field, $h_\|^{\rm intra}$, is
inversely proportional to $\Gamma$ for all hole densities as it is
seen in Fig.~\ref{fig6}(a). This agrees with Eq.~(\ref{eq:kubo1}) and
also Eqs.~(\ref{eq:intraw}) and (\ref{eq:intras}) model for the
ferromagnetic Rashba 2DEG. No $h_\bot^{\rm intra}$ component
exists. On the other hand, the interband part ($a\not=b$) 
of Eq.~(\ref{eq:TSP}) contributes both to $h_\|$ and
$h_\bot$ which is shown in Figs.~\ref{fig6}(b) and (c). The former is a
correction to the intraband SOT field and it scales $h_\|^{\rm
inter}\propto \Gamma$ in the weak scattering limit. It tends to
counteract the intraband contribution, as it is the case in the
ferromagnetic Rashba 2DEG described by
Eqs.~(\ref{eq:sofw}) and (\ref{eq:sofs}). The out-of-plane component
$h_\bot^{\rm inter}$ converges to a finite value when $\Gamma$
vanishes, indicating the intrinsic character of this part of the SOT
field. These results are consistent with the analytical solutions
obtained in Eqs.~(\ref{eq:intras})-(\ref{eq:inter2s}) in the
ferromagnetic Rashba 2DEG and weak scattering limit. It is worth noticing
that this dependence on spectral broadening 
holds over a wide range of $\Gamma$ in the case of
intraband contribution [see inset in Fig.~\ref{fig6}(a)], while it
breaks down already for $\Gamma$ equal to few~meV for the interband
contributions.

\subsubsection{Ferromagnetic splitting}

\begin{figure}[tbh]
\begin{center}
\includegraphics[trim=0mm 0mm 0mm 0mm,clip,scale=0.4]{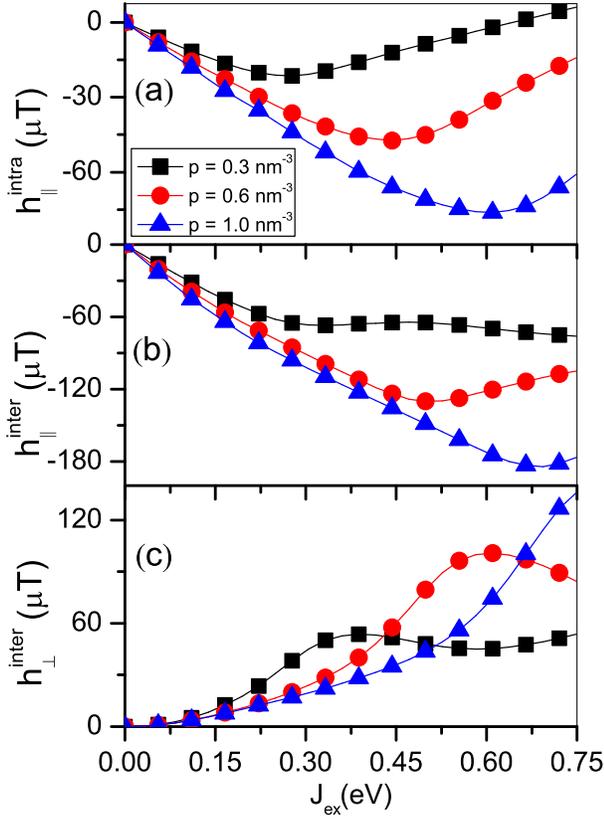}  
\end{center}
\caption{(Color online) (a) Intraband and (b)-(c) interband SOT field
  as a function of exchange interaction $J_{\rm ex}=J_{pd}N_{\tx{Mn}}$.
  Varied values of $J_{\rm ex}$ can be understood as a proxy to
  different Mn doping concentrations, e.g. $x=5\% $ corresponds to
  $J_{\rm ex}=0.06$~eV, the spectral broadening is set to 50~meV and 
  other parameters are the same as in Fig.~\ref{fig6}. }
\label{fig5}
\end{figure}

The band structure of (Ga,Mn)As changes with the Mn doping that would,
in the absence of the SOI, lead to a rigid mutual shift of 
the majority- and minority-spin
bands. Such ferromagnetic splitting would be proportional to 
$J_{\rm ex}=J_{pd}N_{\tx{Mn}}S_{a}$ and we 
can distinguish two limiting situations in a
system where the SOI is present: $E_{SO}\ll J_{\rm ex}$ and
$E_{SO}\gg J_{\rm ex}$. In view of the analytical results presented in
Sec.~\ref{sec:2deg}, it is meaningful to take $E_{SO}=\alpha k_F$ in
the Rashba 2D system. For each component of the non-equilibrium
spin-density $\delta{\bf S}^{\rm intra}$,
$\delta{\bf S}^{\rm inter}_{1}$, $\delta{\bf S}^{\rm inter}_{2}$,
there is a transition between different types of behaviour in the two
limits. For example, the out-of-plane component of the SOT field 
${\bf h}$ changes from the $\propto J_{\rm ex}^2$ behaviour in the
$\alpha k_F\gg J_{\rm ex}$ limit implied by Eq.~(\ref{eq:inter2w}) into a
$J_{\rm ex}$-independent behaviour in the opposite $\alpha
k_F\ll J_{\rm ex}$ limit implied by Eq.~(\ref{eq:inter2s}). We checked
that this transition occurs also in the numerical calculations across
a range of $J_{\rm ex}$ values.

Contrary to the Rashba 2D system, the situation is more complicated in
(Ga,Mn)As because of the additional SOI terms in Eq.~(\ref{eq:Ha5}). 
Due to their mutual competition, it is not obvious what should be taken 
for the effective spin-orbit strength $E_{SO}$.
Looking at the $J_{\rm ex}$-dependence of the individual
SOT field components in Fig.~\ref{fig5}, we nevertheless recognize
similarities to the $E_{SO}\gg J_{\rm ex}$ limit behaviour of the
Rashba 2D system. To some extent, this is a surprising finding since
the disorder broadening used for calculations in Fig.~\ref{fig5} is
quite large ($\Gamma=50$~meV), better corresponding to realistic
(Ga,Mn)As samples but further away from the assumptions used to derive
the analytical results presented in Sec.~\ref{sec:2deg}.  
When $J_{\rm ex}$ is small, both
$h_\parallel^{\mathrm{intra}}$ and $h_\parallel^{\mathrm{inter}}$
are proportional to $J_{\rm ex}$ as seen in Eqs.~(\ref{eq:intraw})
and~(\ref{eq:inter1w}), respectively. On the other hand, 
$h_\perp^{\mathrm{inter}}\propto J_{\rm ex}^2$ in the bottom panel of
Fig.~\ref{fig5} which is reminiscent of
Eq.~(\ref{eq:inter2w}). No similarities to the Rashba 2D system
behaviour of the opposite limit ($E_{SO}\ll J_{\rm ex}$) are found in
our calculations for (Ga,Mn)As.

\subsubsection{Hole concentration}

\begin{figure}[tbh]
\begin{center}
\includegraphics[trim=0mm 0mm 0mm 0mm,clip,scale=0.4]{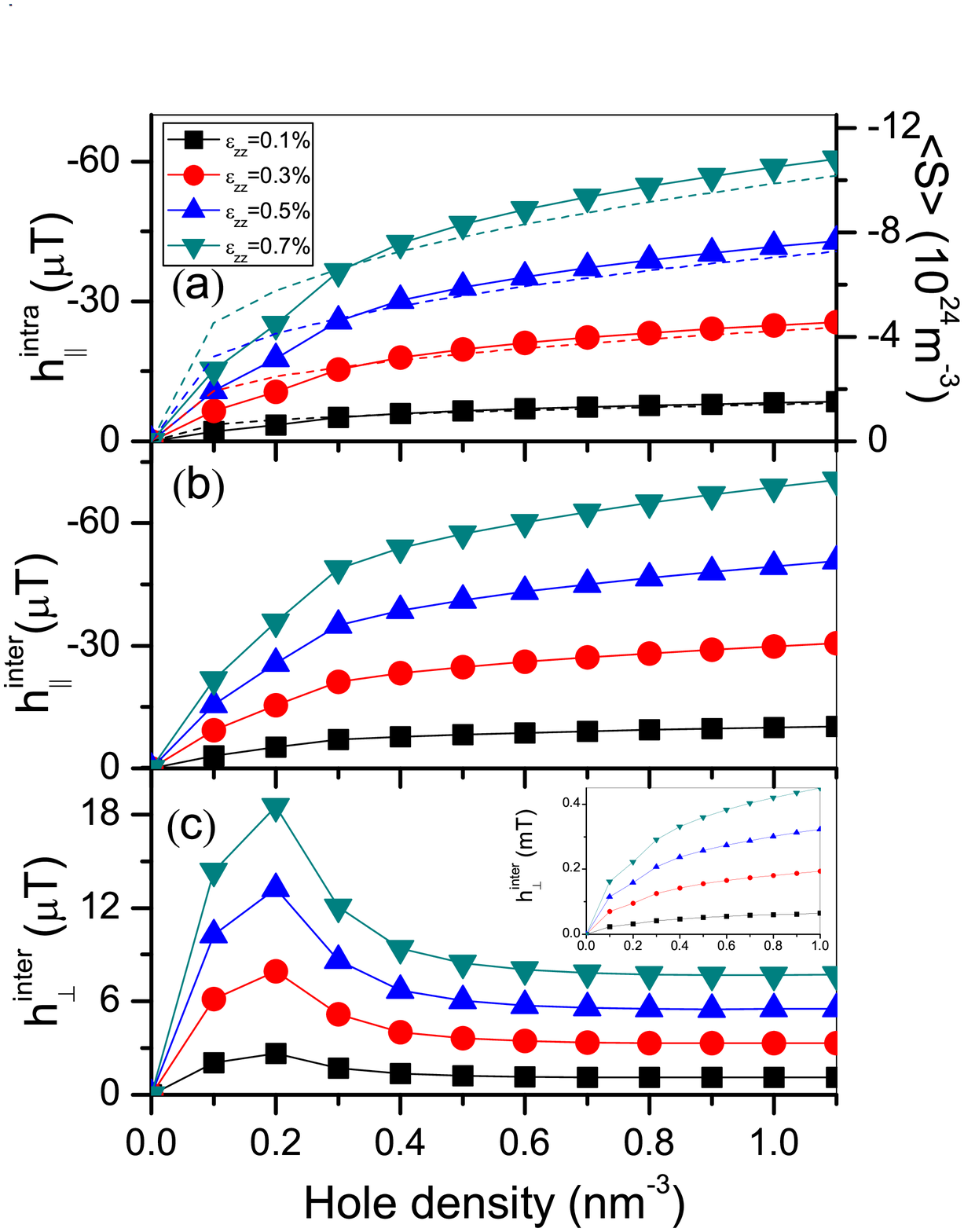}
\end{center}
\caption{(Color online) (a) Intraband and (b)-(c) interband SOT field
  as a function of hole concentration for different lattice-mismatch
  strain $\eps_{zz}$. Inset in (c): interband SOT field in the
  parabolic model. The dashed lines in panel (a) are calculated using
  Eq.~(17) in Ref.~\onlinecite{bernevig} and follow a
  $p^{1/3}$-law. Parameters are the same as in Fig.~\ref{fig5} except
  for $J_{pd}N_{\tx{Mn}}$ fixed to a value corresponding to Mn doping 
  $x=5\% $.}
\label{fig7}
\end{figure}

We display in Fig.~\ref{fig7} the SOT field as a function of the hole
density for different magnitudes of the lattice-mismatch strain
$\eps_{zz}$. First of all, we notice that the SOT field components
increase linearly with the strain. Second, 
increase of the hole concentration results in an
increase in the in-plane SOT field $h_\|$ approximatively
following a $p^{1/3}$ law, as shown in Figs.~\ref{fig7}(a) and (b).
This is consistent with Eq.~(17) in Ref.~\onlinecite{bernevig} in case
of the intraband component. Interestingly, the in-plane interband SOT
field $h_\|^{\rm inter}$ shows a similar tendency
[Fig.~\ref{fig7}(b)], while the out-of-plane interband SOT field
$h_\bot^{\rm inter}$ has a different dependence on $p$.
This anti-damping-like SOT field in Fig.~\ref{fig7}(c) first
increases with the hole concentration in the low hole density regime
and later decreases towards a saturated value. This could be
because of the competition of the different SOI types in (Ga,Mn)As
as noticed by Kurebayashi et al. \cite{Kurebayashi}. Indeed, when the
diamond-lattice spin-orbit coupling is absent ($\gamma_2=\gamma_3=0$), the
out-of-plane interband SOT field $h_\bot^{\rm inter}$ increases with
the hole concentration following the same $p^{1/3}$ law as for the
in-plane field [see inset of Fig.~\ref{fig7}(c)]. For a four-band
Luttinger model that includes band warping ($\gamma_2\neq\gamma_3$), 
the competition between the diamond spin-orbit coupling and the
strain-induced spin-orbit coupling results in a reduction of
$h_\bot^{\rm inter}$, as shown in Fig.~\ref{fig7} of Ref.~\onlinecite{Kurebayashi}. The
reason why the competition between the diamond spin-orbit coupling and
the strain-induced spin-orbit coupling leads to the deviation from the
analytical formula only in the case of the $h^{\rm inter}_\perp$ and
not for $h^{\rm intra,inter}_\|$ remains to be explored in detail.

At this point, we remark that shear strain in Eq.~(\ref{eq:Ha}) leads
to $h^{\rm intra}_\|$ comparable to values shown in Fig.~\ref{fig7}(a) when
the value of $\eps_{xy}$ is comparable to $\eps_{zz}$ used in Fig.~\ref{fig7}.
However, since the relevant values of $\eps_{xy}$ in unpatterned 
epilayers are typically order-of-magnitude lower\cite{Zemen:2009_a}
than those of $\eps_{zz}$, we can typically expect an
order-of-magnitude smaller $h^{\rm intra}_\|$ originating from the
$C_5$-term in Eq.~(\ref{eq:Ha}) as compared to the $C_4$-term.

\subsubsection{\label{sec:fermi-surface}Impact of the Band Structure}

The total DMS Hamiltonian given in Eq.~(\ref{eq:Ha_o}) has both centro-symmetric and non-centro-symmetric components given by Eqs.~(\ref{eq:Ha5}) and (\ref{eq:Ha}), respectively. As discussed in the previous section, the spin-orbit coupling of the centro-symmetric component of the Hamiltonian [\ie the terms in Eq.~(\ref{eq:Ha5}) proportional to $\gamma_2$ and $\gamma_3$] affects also the SOT field, notably their dependence on the magnetization direction [recall the definition of $\varphi$ and $\theta$ below Eq.~(\ref{eq:total-hamiltonian2})]. Apart from the findings of Ref.~\onlinecite{Kurebayashi} discussed above, it was shown in Ref.~\onlinecite{Hang-apl-2013} that the shape of the Fermi surface has a strong impact on the angular dependence of the intraband SOT field $h_\|^{\rm intra}$.    

\begin{figure}[tbh]
\centering
\includegraphics[trim = 0mm 0mm 0mm 0mm, clip, scale=0.4]{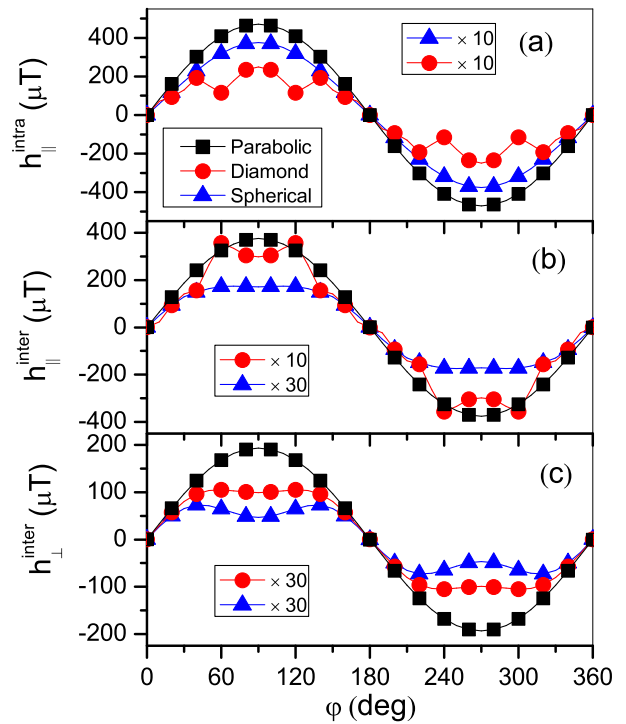}
\caption{(Color online) Intraband and interband SOT field as a
  function of the magnetization direction for different models labelled 
  (i), (ii) and (iii) in the text. The
  red ($\bigcirc$), blue ($\triangle$) and black ($\square$) data stand
  for the full four-band Luttinger model, its spherical approximation
  and the parabolic model, respectively. The parameters are the same
  as in Fig.~\ref{fig7} except for fixed $p=1.0~\tx{nm}^{-3}$ and 
  $\eps_{zz}=-0.3\%$.}
\label{fig1}
\end{figure}

We now systematically explore the influence of the spin-orbit coupling of the diamond crystal on the different components of the SOT field, i.e. $h^{\rm intra}_\|$, $h^{\rm inter}_\|$ and $h^{\rm inter}_\perp$. The centro-symmetric component of the total DMS Hamiltonian, Eq.~(\ref{eq:Ha5}), accounts for the spin-orbit coupling through a set of the Luttinger parameters, $\gamma_{1,2,3}$. By tuning these three parameters, one can modify the form of the centro-symmetric spin-orbit coupling. We model three distinct cases: (i) the parabolic approximation where no centro-symmetric spin-orbit coupling is present ($\gamma_1=2.0,\gamma_2=\gamma_3=0$), (ii) the spherical approximation where the centro-symmetric spin-orbit coupling is turned on but spherical symmetry is retained ($\gamma_2=\gamma_3=2.5$) and (iii) the diamond crystal where both cubic symmetry and centro-symmetric spin-orbit coupling are accounted for  ($\gamma_2\neq\gamma_3$). This approach allows us to identify the role of the last two terms of Eq.~(\ref{eq:Ha5}) on the SOT fields. In Fig.~\ref{fig1}, we show the angular dependence of the different contributions to the SOT field for the spin-orbit coupling induced by the lattice-mismatch strain in the context of models (i)--(iii). The magnetization lies in the ($x,y$) plane ($\theta=\pi/2$) and its direction is given by the azimuthal angle $\varphi$.

As expected from the symmetry of the $C_4$ term in Eq.~(\ref{eq:Ha}), the three components of the SOT field have a dependence of the form $\sin\varphi$ in the parabolic model ($\square$ symbols in Fig.~\ref{fig1}). When diamond-lattice spin-orbit coupling is switched on but the spherical approximation is assumed, the interband SOT fields [$\triangle$ symbols in Figs.~\ref{fig1}(b) and (c)] deviate from this dependence, while the angular dependence of the intraband term remains unaffected [$\triangle$ symbols in Fig.~\ref{fig1}(a)]. Furthermore, the magnitudes of interband and intraband SOT fields strongly decrease. This is a manifestation of the competition between the strain-induced terms in Eq.~(\ref{eq:Ha}) with the centro-symmetric Luttinger spin-orbit terms in Eq.~(\ref{eq:Ha5}).\cite{Kurebayashi}\par

When the spherical approximation is lifted ($\gm_{2}\neq \gm_{3}$)
electronic bands become warped, especially those of the heavy holes. This
results into an increase of the interband SOT fields and an additional
angular dependence shown by $\bigcirc$~symbols in
Fig.~\ref{fig1}. Microscopically, the latter effect is caused by the
distorted spin textures on the Fermi surface. The influence of the
centro-symmetric spin-orbit field on the spin torque in GaMnAs has
also been identified by Haney~et~al.\cite{haneyprl} in DMS
spin-valves.

\section{\label{sec:sum}Summary and conclusion}

We have studied the intraband and interband SOT fields using Kubo
formula, in the prototypical case of a ferromagnetic 2DEG with Rashba
spin-orbit coupling, as well as in a three-dimensional DMS modelled by
a kinetic-exchange Kohn-Luttinger Hamiltonian. For the latter, parameters
pertaining to (Ga,Mn)As were used. In the limit of low doping
concentration and weak exchange coupling, we find similarities
between the two systems, demonstrating that the general
trends of the intrinsic and extrinsic SOT fields can be understood
analytically using the Rashba 2DEG in the weak scattering
limit. Nevertheless, the numerical analysis of the three-dimensional
DMS system also unravels the complex interplay between the
different types of spin-orbit coupling (centro-symmetric and 
non-centro-symmetric) involved in realistic systems resulting in
complex dependences of the SOT fields on the magnetization direction 
as well as significant differences from the
Rashba 2DEG model. The contribution of interband mixing to the SOT
presents an outstanding opportunity to explain the emergence of large
anti-damping-like torques that cannot be readily attributed to spin
Hall effect, offering an interesting platform to interpret recent
puzzling results.\cite{jamali,mellnik}

\acknowledgements

I.G. acknowledges financial support from Canadian NSERC. H.L., X.W.,
F.D., and A.M.  were supported by the King Abdullah University of
Science and Technology (KAUST). H.G. and J.S. were supported by the NSF Grant No. DMR-1105512, and the Alexander Von Humboldt
Foundation. Work at FZU was supported by the Grant
Agency of the Czech Republic Grant No. 14-37427G and K.V. moreover 
acknowledges sustained support of L., Grant No. 15-13436S and
Ministry of Education of the Czech Republic Grant No. LM2011026.
We thank J. \v Zelezn\'y for his inspirational suggestions and
A.H.~MacDonald for discussions at initial stages of the project.

\end{document}